\documentstyle[12pt,epsfig]{article}
\def\beq{\begin{equation}}
\def\eeq#1{\label{#1}\end{equation}}
\def\eeqn{\end{equation}}

\def\beqa{\begin{eqnarray}}
\def\eeqa#1{\label{#1}\end{eqnarray}}
\def\eeqan{\end{eqnarray}}

\let\bar=\overbar

\def\Dslash{\not{\hbox{\kern-4pt $D$}}}
\def\dslash{\not{\hbox{\kern-2pt $\del$}}}

\def\msb{{\bar{\ssstyle M \kern -1pt S}}}

\def\Title#1{\begin{center}{\Large #1}\end{center}}
\begin{document}
\Title{Status of KEKB accelerator and detector, BELLE}
\bigskip\bigskip
\begin{raggedright}
{\it F. Takasaki\index{Takasaki F.}\\
Institute of Particle and Nuclear Studies\\
High Energy Accelerator Research Organization \\
Oho 1-1, Tsukuba-shi, Ibaraki-ken 305, Japan}
\bigskip\bigskip
\end{raggedright}

\section{Introduction}
\begin{figure}
\begin{center}
\hspace*{-8mm}
\hbox{
      \epsfig {file=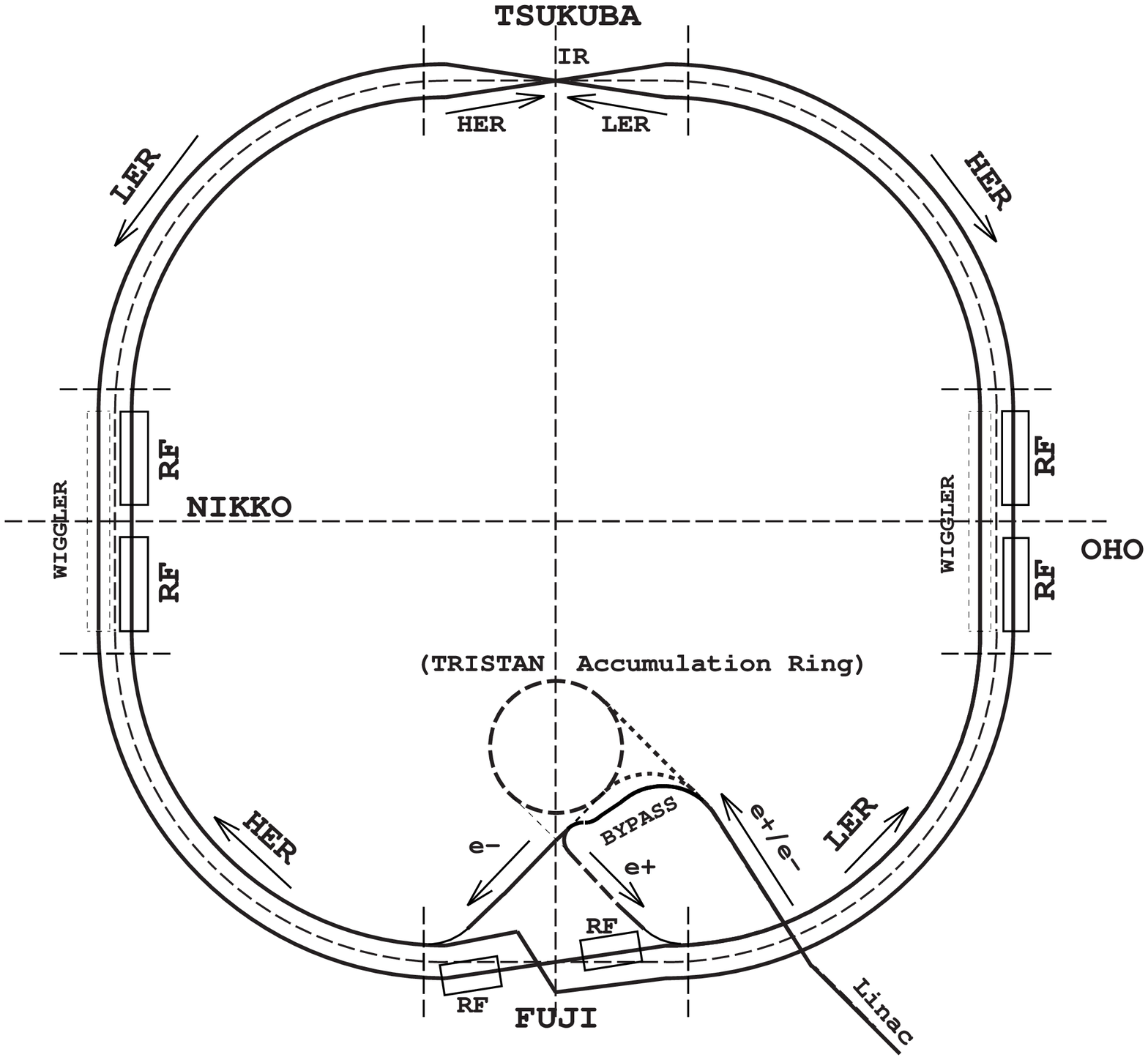 ,width=7.0cm ,height=6.6cm ,angle=0}
      \epsfig {file=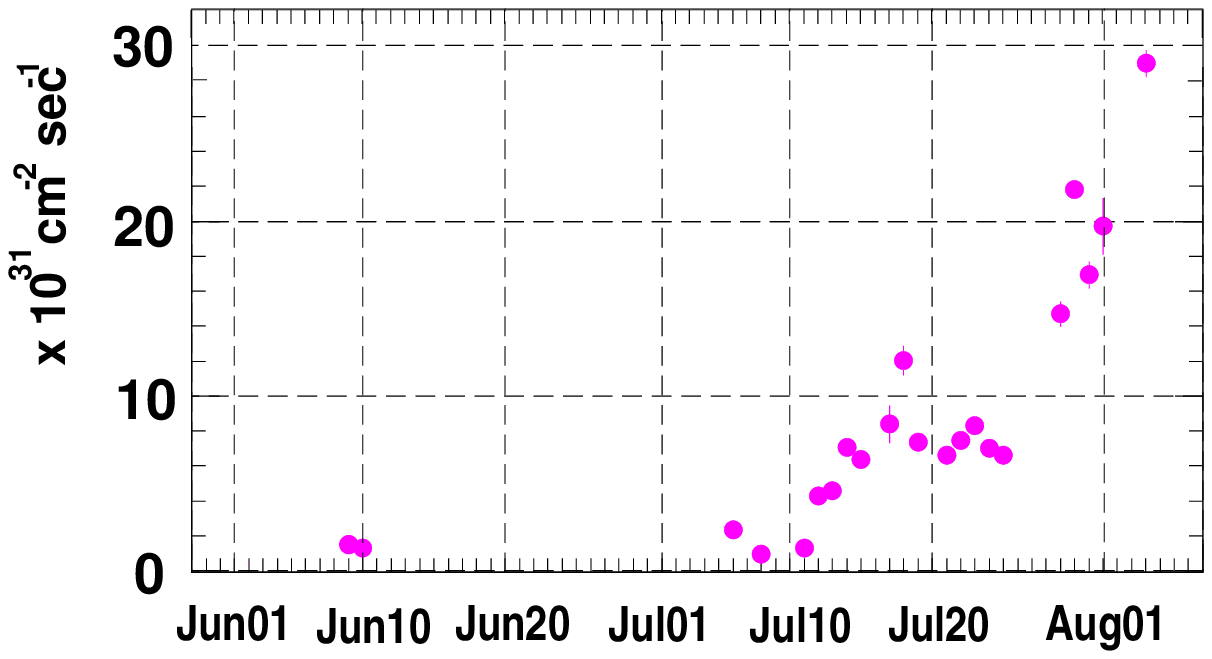 ,width=8.cm ,height=4.5cm ,angle=0}
     } 
\caption{KEKB accelerator complex and history of Luminosity Improvement}
\label{fig:KEKBmap}
\end{center}
\end{figure}

The KEK B-Factory, KEKB, is an asymmetric energy collider with 8 GeV electron and 3.5 GeV positron beams built in the 3~km long TRISTAN tunnel of KEK (Figure~\ref{fig:KEKBmap}). One of the physics goals of this facility is to make a detailed study of the B-meson, in particular to observe possible CP violation effects in its decay.  
The energy difference between the electron and positron beams gives a boost to the produced B-meson pairs. This makes it possible to measure time dependent features of B-meson decay where a large CP asymmetry may show up as predicted by A.~Carter and I.~Sanda~\cite{Sanda}.  The KEKB project was first discussed at KEK in 1987 and the first conceptual design of the accelerator was worked out by K.~Oide in February 1988~\cite{Oide}. It was later modified to the present design with the equal circumference of two rings in 1989~\cite{Hirata}.  The project was approved by the Japanese government in April 1994.

\section{Accelerator design}
The biggest challenge of the project is how to produce the more than 10 million B-meson pairs required to observe CP asymmetry, or how to reach a luminosity greater than 10$^{34}$cm$^{-2}$sec$^{-1}$.

This high luminosity is to be achieved by the collision of beams stored in two  different synchrotrons with equal circumference. We store as many beam bunches as possible in the synchrotrons in which one particular beam bunch in one ring collides with a particular bunch in the other ring.  The single bunch current and therefore the single bunch luminosity is a moderate one which can be easily realized by established technology.  However, the total luminosity which is a sum of many single bunch luminosity can be made quite high~\cite{KEKBdesign}. The designed total beam currents are 2.6~A for the positron (LER) and 1.1~A for the electron (HER) rings.  

A big concern was how to overcome beam instabilities which may show up when storing a large number of bunches and a high beam currents in the synchrotrons.  There are instabilities due to single-beam collective effects, impedance from various beamline elements, ion trapping, photo-electron effects, and beam-beam effects.  These instabilities are suppressed by carefully designing accelerator components such as RF cavities with HOM damping and the smooth surface of accelerator elements as well as by the implementation of powerful feedback systems~\cite{KEKBdesign}. 

The choice of a finite beam crossing angle in the horizontal plane at the beam collision point, 22~mrad, is a notable feature of the accelerator design.  This scheme eliminates parastic collisions and make it much easier to manipulate two beams of different energies around the collision point. It hopefully leads to smaller beam induced backgrounds.  The positron beam enters the solenoid field volume parallel to the field axis, while the incoming electron beam orbit is at an angle of 22~mrad from the field axis.  The integrated field strength along the beam path is locally cancelled by a pair of additional superconducting solenoids with smaller radius placed in the detector solenoid. Extensive simulation studies indicate that the instability due to the finite angle collision of beams can be avoided by choosing the proper sets of betatron tunes for the beams~\cite{KEKBdesign}.  A crab cavity is being developed to rotate the beams to provide head-on collision in case the finite angle crossing causes a serious problem~\cite{KEKBdesign}. 

  Many wiggler magnets are used to reduce the damping time of the positron beam.  The LINAC was upgraded in energy from 2.5~GeV to 8.5~GeV to make it possible to inject electron and positron beam directly into the storage rings~\cite{KEKBLINAC}.  The beam optics of the storage rings remains unchanged both at the time of beam injection and beam storage. The two rings are equipped with more than 1000 beam position monitors to facilitate quick beam diagnostics.

\section{Accelerator commissioning}
The upgrading of the LINAC was completed successfully in July 1998.  The KEKB synchrotrons received the first beam from the LINAC in December 1998.  The commissioning continued until mid April 1999 without BELLE.  Although there have been several interuptions by unexpected accidents, we have succeeded in storing both electron and positron beams larger than 500~mA. The integrated stored current reached 100 A-hour for the positron beam and 70 A-hour for the electron beam~\cite{KEKBCommission}.  The BELLE was rolled onto the beamline in May 1999.

  The collision of the beams was confirmed on June 1 by observing hadronic events with the BELLE detector.  Although interrupted again by an accident for two weeks, the operation of the accelerator continued until August 4th.  
Part of the operation time was dedicated to physics runs.  The total integrated luminosity accumulated by the BELLE detector was 25~pb$^{-1}$.  The highest luminosity averaged over one hour reached 2.5~pb$^{-1}$.  The highest peak luminosity, 2.9$\times$10$^{32}$ cm$^{-2}$sec$^{-1}$ (Figure~\ref{fig:KEKBmap}), was recorded on the last day of the summer run.

The parameters with which the storage rings produced the highest peak luminosity are summarized in Table~\ref{tab:performance}.  

\begin{table}[htb]
\begin{center}
\begin{tabular}{|c|c|c|} \hline
 & LER & HER \\ \hline
Best Peak Luminosity & \multicolumn{2}{c|}{2.9$\times$10$^{32}$cm$^{-2}$sec$^{-1}$} \\ \hline
\# of Bunches/Bunch Trains/Spacing & \multicolumn{2}{c|}{1160/29/3} \\ \hline
Beam Current & 293 mA & 190 mA \\ \hline
Bunch Current & 0.25 mA & 0.16 mA \\ \hline
Beam Size at IP (x) & 190 $\mu$m & 190 $\mu$m \\ \hline
Beam Size at IP (y) & 3.9 $\mu$m & 3.3 $\mu$m \\ \hline
Beta Functions at IP (x/y) & \multicolumn{2}{c|}{1 m/ 0.01 m} \\ \hline
Emittance ratio & 8.5 & 6.1 \\ \hline
Tune Shift (x) & 0.034 & 0.023 \\ \hline
Tune Shift (y) & 0.014 & 0.008 \\ \hline
Life Time  & 100 min@290 mA & 300 min@190 mA \\ \hline
\end{tabular}
\caption{ KEKB accelerator performance}
\label{tab:performance}
\end{center}
\end{table}

Several remarks are in order here.  1) All accelerator components, such as newly developed RF cavities, ARES, single cell superconducting RF cavities, superconducting quadrupole magnets at the IR, and many wiggler magnets, have functioned as designed.  2) The single bunch currents stored are close to the design values.  3) The number of bunches stored is about one fifth of the design values.  For the electron ring the limitation is mainly due to a fast ion trapping effect, while for the positron ring no explanation has yet been determined.  4) The beam sizes were measured by the synchrotron light using an interferometer.  The vertical and horizontal beam sizes were about 3.3~$\mu$m and 190~$\mu$m for the electron beam at 100~mA and about 3.6~$\mu$m and 190~$\mu$m for the positron at 200~mA.  These are larger than the designed values by a factor of 2.  5) The beam blows up vertically as the amount of stored beam currents increase.  This is suppressed by a feed-back system quite effectively for the electron beam but not for the positron beam.  6) The beams blow up vertically to about 6~$\mu$m when the beams collide at 100~mA (HER) and 200~mA (LER). 
  
Although the luminosity achieved is only 3\% of the design value and we have many things to be understood, we believe that the basic design concept, especially the choice of the finite angle crossing scheme, is not wrong.   The factor of 30 improvement will be achieved by a careful tuning of the accelerator parameters.  

\section{BELLE commissioning}
BELLE \cite{BELLEdesign} is a solenoid spectrometer with the capability of precise energy and momentum measurement and of good particle identification, especially $\pi$-K separation.  The field strength of the solenoid is 1.5 Tesla with the field uniformity better than 4\% in the central tracking volume.  The detector is arranged asymmetrically along the direction of the boost.  The detector has been built by the collaboration of 51 institutions from ten countries and one region \cite{BELLE}.

The construction began in April 1994 and was completed in December 1998.  Data taking began in June 1999 with all sub-detectors assembled.  Data taking was made mostly on the peak of the Upsilon(4s) resonance after its confirmation by an energy scan.  
The integrated luminosity recorded by the detector was only 25pb$^{-1}$ by August 4th. Therefore, no physics results can be presented in this report but rather we report on the detector performance obtained by using about 60,000 hadronic events.  Although we suffered from beam induced backgrounds, data taking was successful.  The typical trigger rate was about 300~Hz at the beam currents of 200~mA for LER and 100~mA for HER.   The analysis of the data has proceeded smoothly using software developed in the last five years.   
 
\subsection{Vertexing and tracking} 
The BELLE vertex detector(SVD) is a three layer double sided silicon strip detectors and the tracker is a standard multi-cell drift chamber.  The SVD covers the polar angle between 20$^\circ$ and 140$^\circ$ in the lab frame.  The impact parameter resolutions are measured to be 35~$\mu$m and 40~$\mu$m in r-$\phi$ and z coordinates, respectively.  The momentum resolution is about 0.5\% at 1~GeV. The following examples illustrate how the vertexing and tracking work.

\begin{figure}
\begin{center}
\vspace*{-8.5cm}
\hspace*{-60mm}
\hbox{
    \epsfig {file=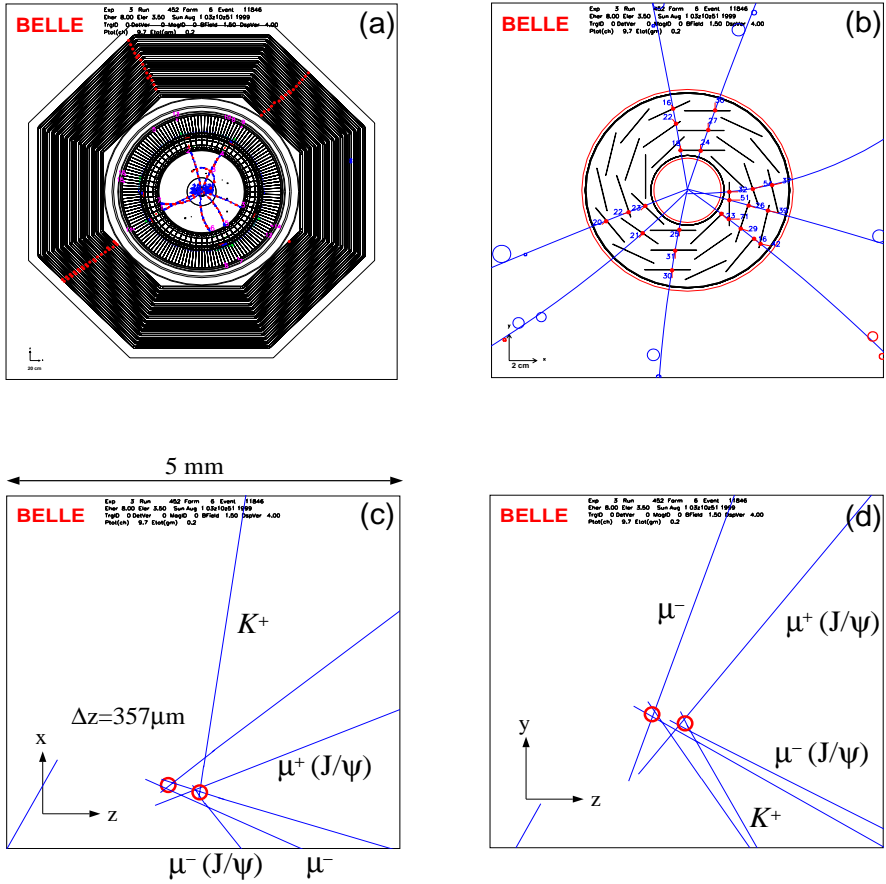 ,width=14.cm ,height=12.5cm ,angle=0}
    } 
\vspace*{7.6cm}
\caption{
(a)An event of one B-meson decaying to J/$\psi$ and K$^+$ (b)the other B-meson decaying into a state including a muon and its hit pattern in the silicon vertex detector
(c)(d)A zoom up view of the same event along the beam direction
}
\label{fig:BJK}
\end{center}
\end{figure}

First, we show a candidate event where one B-meson decays to J/$\psi$ and K$^+$ and the other B-meson decays to a state including a $\mu$$^-$.  Two vertices are separated by 357~$\mu$m along the z direction, which is substantially larger than the vertex resolution.  It should be noted that the $\mu$$^\pm$ and K$^+$ are positively identified (Figure~\ref{fig:BJK}).

\begin{figure}
\begin{center}
\vspace*{-5mm}
\hspace*{0mm}
\hbox{
    \epsfig {file=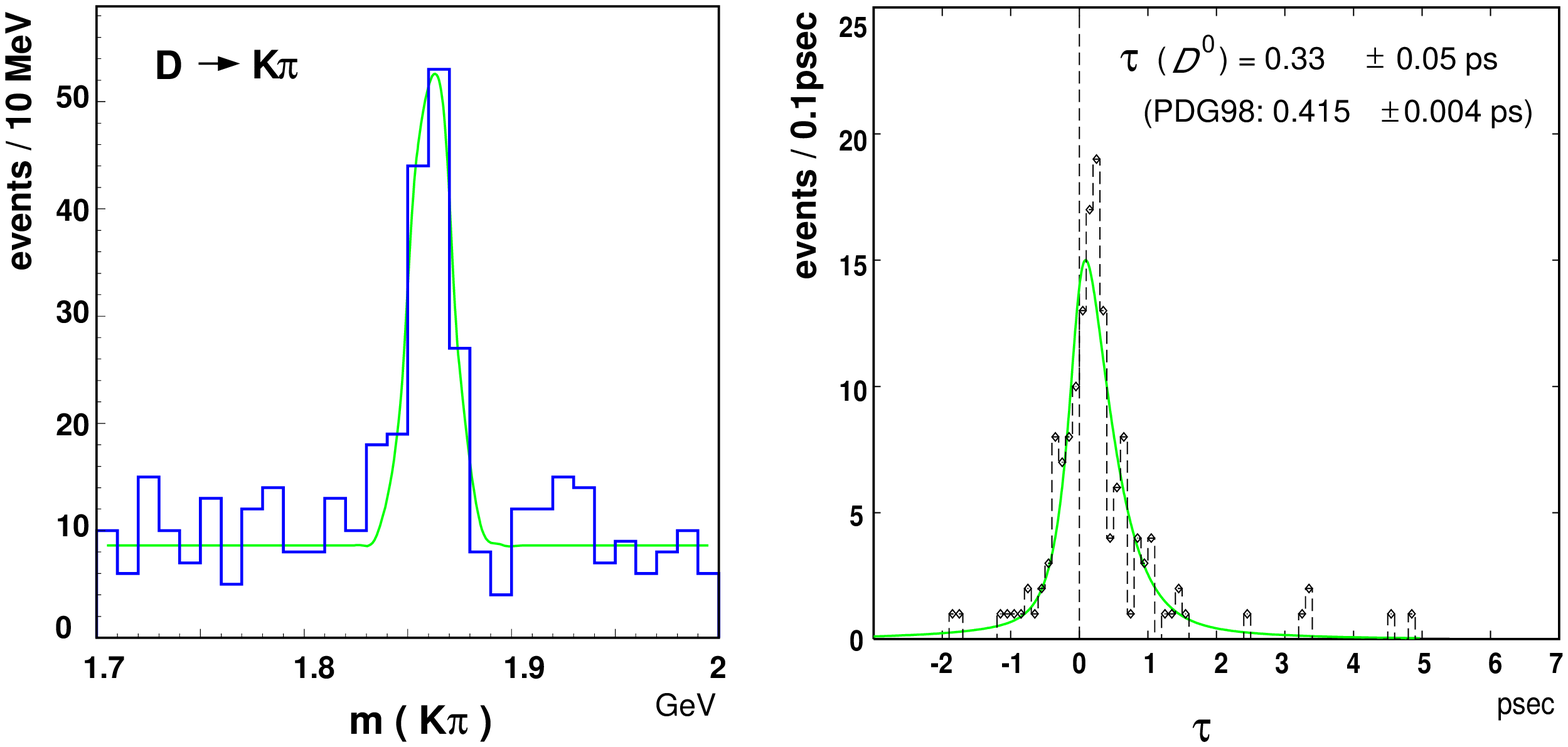    ,width=13.cm ,height=6.cm ,angle=0}
    } 
\caption{Invariant masses for D$\rightarrow$K$\pi$ and its decay length distribution.}
\label{fig:Dlife}
\end{center}
\end{figure}

Another example of the BELLE vertexing performance is the measurement of the lifetime of D-meson.  In this study, we have chosen D$\rightarrow$K$\pi$ samples with momentum larger than 2.4~GeV.  The distance between the beam center and the vertex point in the plane transverse to the beam direction is plotted in the figure~\ref{fig:Dlife} together with an invariant mass distribution for D$\rightarrow$K$\pi$. The lifetime measured is consistent with the known value.  

\begin{figure}
\begin{center}
\vspace*{-0.5cm}
\hspace*{0mm}
\hbox{
    \epsfig {file=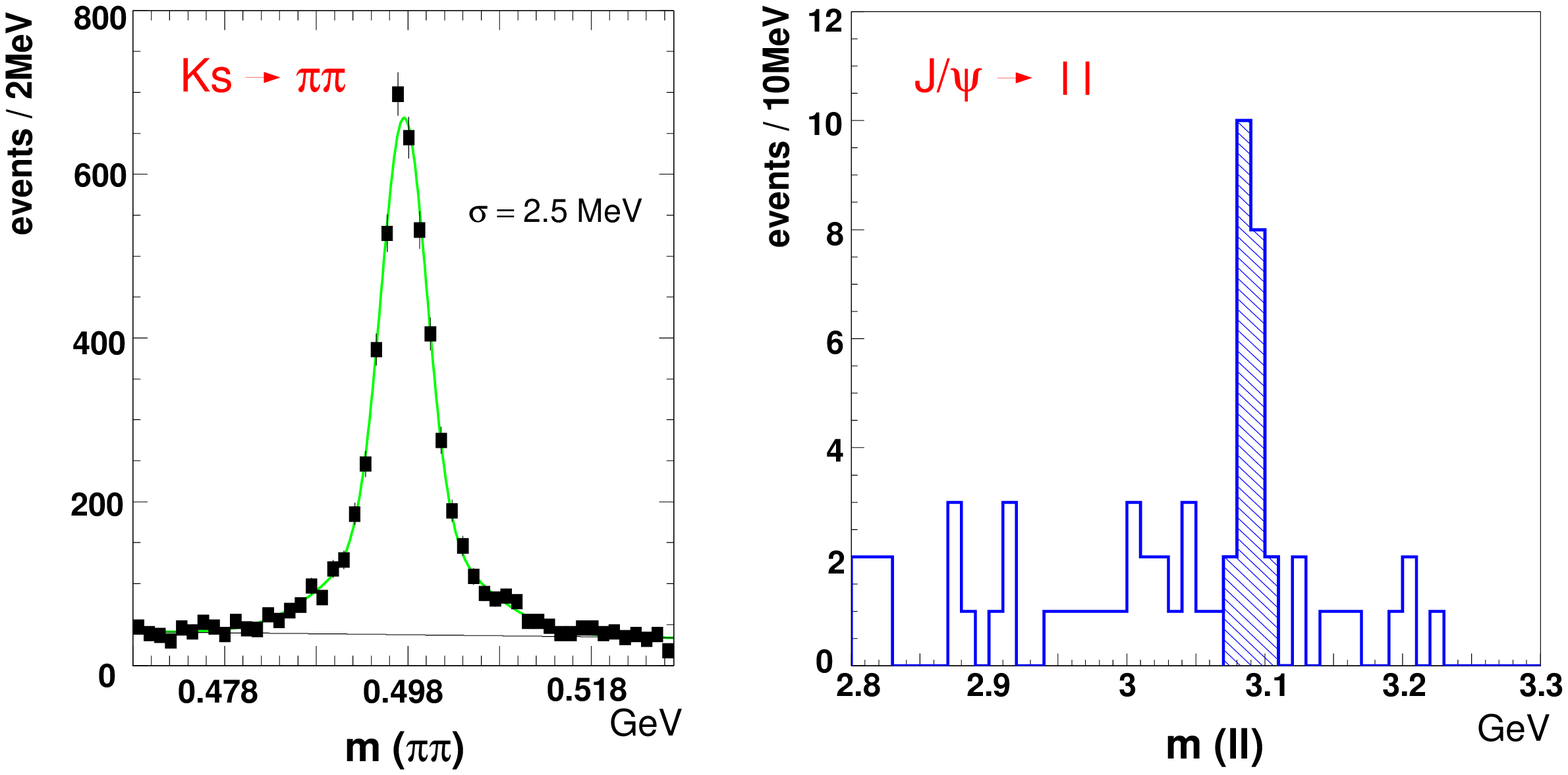   ,width=13.cm ,height=6.cm ,angle=0}
    } 
\caption{Invariant masses for Ks decaying to $\pi$$^+$-$\pi$$^-$
and J/ $\psi$ decaying to two leptons }
\label{fig:Ks}
\end{center}
\end{figure}

The next examples are the invariant mass distributions for Ks$\rightarrow$$\pi$$^+$$\pi$$^-$ and J/$\psi$$\rightarrow$${\sl l}$$^+$${\sl l}$$^-$ observed in hadronic events (Figure~\ref{fig:Ks}). They give an idea of the performance of tracking.   

\subsection {$\pi$-K separation}  
\begin{figure}
\begin{center}
\hspace*{-1mm}
\hbox{
    \epsfig {file=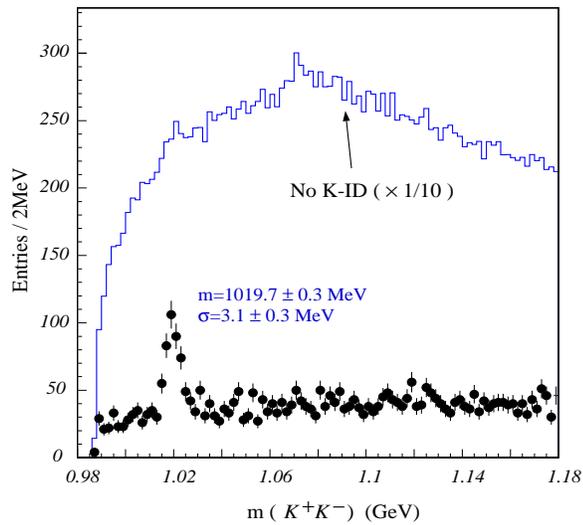 ,width=8.cm ,height=7.cm ,angle=0}
    } 
\caption{Invariant masses calculated for two charged particles assumed to be K-mesons(histogram) and those that passed the K-meson requirements (Black circle)}
\label{fig:phi}
\end{center}
\end{figure}

The BELLE detector is equipped with three different types of detectors for charged K-meson identification: The aerogel Cherenkov counters, the time of flight counters and the drift chamber.  We calculate the K-meson likelihood for all charged tracks by combining the information from the three particle id devices. The K-meson identification is made by selecting tracks with high K-likeliness.  To show how this method works, we present a plot of invariant mass made for two tracks with opposite charges for all possible combination in hadronic events.  The histogram is the invariant mass assuming two tracks to be K-mesons, while the black circles are those requiring both tracks to have K-meson probabilities greater than 70\%.  One clearly sees a peak corresponding to the $\phi$-meson with a background suppression of more than a factor of 100 (Figure~\ref{fig:phi}). It has been confirmed that this method is also quite effective for the identification of charged K-mesons in the decay process D$\rightarrow$K$\pi$. 

\subsection {The calorimeter}
\begin{figure}
\begin{center}
\hspace*{-1mm}
\hbox{
    \epsfig {file=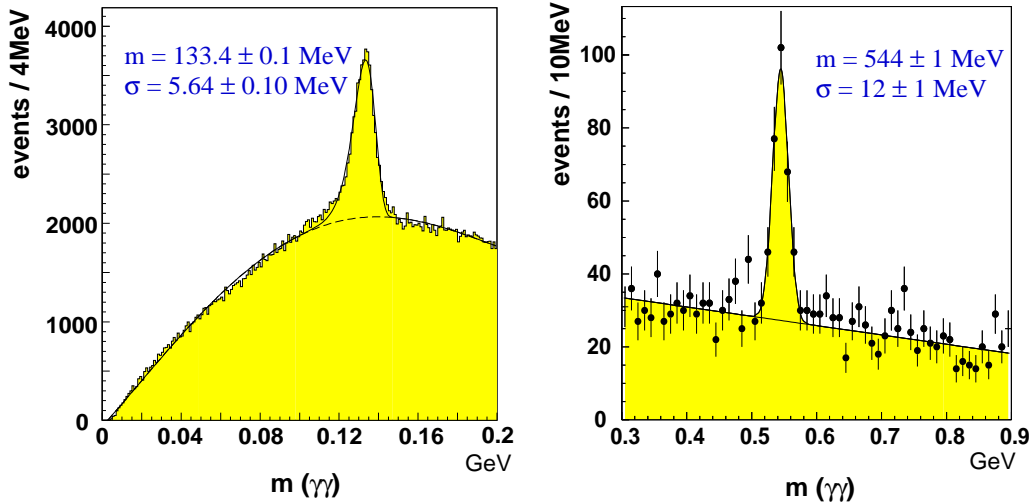 ,width=14.cm ,height=7.0cm ,angle=0}
    } 
\caption{Invariant mass for two gamma rays with energy larger than 50 MeV }
\label{fig:pieta}
\end{center}
\end{figure}

The Belle electromagnetic calorimeter is made of finely segmented CsI crystals of about 30~cm in length.  The calibration of the calorimeter has been performed by using cosmic rays and Bhabha events. We present an invariant mass spectrum formed for two energy clusters which are greater than 50~MeV in the  figure ~\ref{fig:pieta}.  One sees clear peaks for $\pi$$^\circ$ and $\eta$ mesons.

 \section{The beam background}
  In spite of the choice of the finite angle beam crossing scheme at the interaction point, we have suffered from beam induced backgrounds. Despite the backgrounds, data taking was possible with the largest stored beam current in collision mode, about 200~mA for the positron beam and 100~mA for the electron beam.  

  The most serious background was due to synchrotron radiation with a critical energy of a few keV from the electron beam generated at magnets upstream of the detector.  These photons were not expected from simulation nor were they observed by the dedicated beam background detector before BELLE rolling-in. They could be eliminated by adjusting the steering of the electron beam orbit. However, we noticed this too late and this damaged the inner layer of the silicon detector.

  We suffered also from synchrotron radiation with a critical energy around 30 keV.  This background is presumably generated by the outgoing electron beam in the superconducting quadrupole magnet which generates a fan that hits the aluminum part of the beam pipe and then scatters back into the detector.  This synchrotron light will be suppressed by replacing the beam pipe with an appropriate shape made of copper.  

 The BELLE detector was not free from over-focused beam particles, especially from the electron beam, which hit the detector.  This backgrounds will be suppressed by improving the vacuum and by installing shielding around the beam pipe.  

\section{Summary}
The construction and the commissioning of the KEKB accelerator and the detector has been successful.  It was shown that the BELLE detector functions as designed.  The time spent on commissioning before the summer shut-down was so short that we could achieve only 3\% of design luminosity.  The luminosity will be improved by carefully tuning the beam parameters.   During the summer break in KEKB operation, more RF cavities are installed and more shielding around the interaction region is added and the damaged silicon detectors are replaced.  KEKB operation will resume in October and will continue for 10 months.  We hope that we will have information on CP violation in B-meson decay by the end of next run.

\def\Discussion{
\setlength{\parskip}{0.3cm}\setlength{\parindent}{0.0cm}
\bigskip\bigskip{\Large {\bf Discussion}} \bigskip}
\def\speaker#1{{\bf #1:}\ }

\Discussion

\speaker{Richard Taylor (SLAC)}
You mentioned that the data analysis is under control.  Will this still be the case when the luminosity increase by a factor of 30?

\speaker{Takasaki}
We are prepared for much higher rate and we hope it will be OK.


\begin{thebibliography}{99}
\bibitem{Sanda}
A. B. Carter and A. I. Sanda,  Phys. Rev. Lett. 45(1980) 952
\bibitem{Oide}
K. Oide :  Private communication
\bibitem{Hirata}
K. Hirata and E. Keil,  Phys. Lett. B232, 413(1989)
\bibitem{KEKBdesign}
   KEK Report 95 - 7
\bibitem{KEKBLINAC}
   KEK Report 95 - 18
\bibitem{KEKBCommission}
 KEK Preprint 99 - 8
\bibitem{BELLEdesign}
  KEK Report 95 - 1
\bibitem{BELLE}
Academia Sinica, Aomori University, Budker Institute of Nuclear Phyics, Chiba University, Chuo University, University of Cincinatti, Institute of Cosmic Ray Reseaches(University of Tokyo), Fukui University, GyeonSang National University, University of Hawaii, Institute of High Energy Physics, Joint Crystal Collaboration Group, Kanagawa University, KEK, Korea University, Krakow Institute of Nuclear Physics, Kyoto University, Melbourne University, Mindanao State University, Nagasaki Insitute of Applied Science, Nagoya University, Nara Women's University, National Central University, National Kaoshin Normal University, National Lien Ho C.T.C., National Taiwan University, Nihon Dental College, Niigata University, Osaka University, Osaka City University, Princeton University, Saga University, SankyunKwan University, University of Science and Technology of China, Seoul National University, Institue of Single Crystal, Sugiyama Jogakuin University, University of Sydney, Institute of Theoretical and Experimental Physics, Toho Univeristy, Tohoku University, Tohoku-Gakuin University, University of Tokyo, Tokyo Institute of Technology, Tokyo Metropolitan University, Tokyo University of Agricalture and Technology, Toyama National College of Maritime Technology, University of Tsukuba, Utkal University, Yonsei University, and Virgina Polytechnic Institute      
\end{thebibliography}
\end{document}